# Coordination Strategies When Working from Anywhere: A Case Study of Two Agile Teams

Tor Sporsem[1][0000-0002-5230-7480] and Nils Brede Moe[1][0000-0003-2669-0778]

[1] SINTEF Digital, 7034 Trondheim, Norway
`tor.sporsem@sintef.no`

**Abstract.** Effective coordination is the key to successful agile teams. They rely on frequent interactions and mutual adjustment to manage dependencies between activities, which traditionally has been solved by co-locating the team. As the world is adjusting to post-covid work-life, companies are moving towards a work-from-anywhere approach where workers can choose to what degree they want to work from home or office. However, little is known about coordination in such a context. We report findings on developers' emerging strategies when working-from-anywhere, from an exploratory case study in Norway, including eight interviews. Our study shows that new strategies for mutual adjustment emerged as teams experimented with different tools and approaches: developers chose tasks according to location, tasks with vague requirements are performed collocated while individual tasks requiring focus are best performed at home; large meetings are virtual, preserving co-located time for collaborative tasks; using virtual rooms to maintain unscheduled meetings as they communicate mental presence to teammates, lowering the threshold for intra-team unscheduled talks. The strategies can help organizations create a productive and effective environment for developers.

**Keywords:** WFX, work from home Large-scale agile coordination, co-located, mutual adjustment, unscheduled meetings, virtual rooms, Discord, Slack, hybrid

## 1 Introduction

In March 2020, technology companies closed their offices and sent employees to work from home (WFH), due to the Covid-19 pandemic. While some reported a decrease in developer productivity a recent study [1] found that many software developers benefit from WFH, and argued that most developers do not want to fully return to the office, while at the same time teamwork suffers. Therefore, many companies will opt for a hybrid workplace – office days mixed with WFH days. Consequently, companies like Facebook, Twitter, Square, Shopify, and Slack have established policies of long-term or even permanent working from home [2]. Spotify announced the Work-from-Anywhere (WFX) policy that allows employees to choose how often they prefer to be in the office or at home, or somewhere else. At the same time, there is little knowledge about the long term effects of WFX. We have little knowledge on consequences for learning, coordination and solving tasks [1].

In agile teams, work relies heavily on coordination by feedback and mutual adjustment, particularly in meetings and ad hoc conversations [3]. Therefore, distributed agile teams need an effective coordination structure, with both scheduled and unscheduled meetings and the right informal collaboration tools to support mutual adjustment [4]. However, mutual adjustment in its pure form requires everyone to communicate with everyone [5]. Coordination by mutual adjustment is challenging when part of the team is working full time from home or from the office, or the whole team is working from anywhere. Also, it is challenging to know what collaboration should occur when the team is co-located, which sometimes is only a few times per week, month, or year. Given that coordination by mutual adjustment is essential for agile teams, and that more and more organizations are implementing practices for working from anywhere, we identified the following research question: *What coordination strategies are used by agile teams when working from anywhere?*

To answer, we report empirical insight from a case study on two developer teams in the company Entur. Since the study is exploratory, we have included both inter- and intra-team coordination. Section II describes related work. Section III outlines our research method and case context, followed by our findings. Section V discusses the strategies found and compares them to related research, concludes our work, and points to future research.

## 2    Coordinating work in distributed agile teams

Agile practices have stretched from the intended ideal of small co-located teams and reached safety-critical, large-scale, and distributed software development programs. Effective coordination is the key to success for agile teams in all contexts. A key to coordination "is managing dependencies between activities" [6]. In agile teams, coordination is exercised through several mechanisms [7]. As agile software development relies on frequent interactions and mutual adjustment, and since physical distance makes people communicate less [8], virtual teams need tools that can mitigate the barriers of distance and reduced communication.

In their study of distributed teams, Stray and Moe [4] found the IM tool Slack to be one of the most important collaboration and coordination tools. While Slack supported coordination in the distributed teams, the research by Stray shows that some users were very active, while others posted very few messages. Further, experienced team members favored messages in open channels while less experienced people favored more direct messages (i.e.one-to- one communication). At the same time, Slack causes interruptions. In their study of a globally distributed project, Matthiesen et al. [9] found that interruptions on IM tools were perceived as normal or as negative disruptions, depending on the quality of the relationships between the distributed colleagues. While tools are important, Calefato et al. [10] argue that face-to-face meetings are essential for having more in-depth discussions. In line with this, Stray [4] found the importance of co-locating permanent distributed teams once or twice a year and that the most complex and challenging meetings be organized during the co-location periods. In global software development, the setup is planned and voluntarily. In March 2020 most had to go home. To understand how WFX can work, there is a need to understand what happened during the pandemic, and specially why some teams struggled.

During the pandemic, several explanations have been found for why developers and teams had problems managing dependencies between team members. Examples are connectivity problem and poor workspace equipment , lack of match of working hours in the team, and greater difficulty in interpersonal communication [11, 12]. Smite et al. [1] found a reduced speed of solving tasks resulting from an increased number of meetings, worse understanding of what is going on in the team, and exhaustion from running meetings virtually. Furthermore, brainstorming sessions and problem-solving sessions were reported to be more challenging and to require more time due to the lack of accustomed whiteboards, possibility to spontaneously connect to the needed people, and requiring considerably more time to prepare. Finally, developers have stopped pair programming practices because they lack tool support or are not aware of the status of other team members [13]. At the same time, many have reported more effective task solving and work coordination from the home office. Reasons include better focus time, fewer interruptions, more time to complete work, more efficient meetings , and a better/more comfortable work environment [11, 12]. Smite et al. [1] found fewer distractions and interruptions, increased flexibility to organize ones work hours, and easier access to developers a person depend on to complete the work. While tasks are solved more effectively, coordination suffers [1].

## 3   Method

To answer the research question, we conducted a case study, investigating practices in two developer teams at Entur; a public, mature large-scale agile development company. We chose this case because Entur is part of an established research program. Entur has twenty development teams, and each team is responsible for their part of the digital infrastructure they deliver to the Norwegian public transport system. Prior to Covid-19, the teams used tools such as Slack, Jira, and Confluence, and material artefacts such as task boards. The teams chose freely how they go about solving their tasks and rely on agile methods of choice. As such, there was no one unified agile approach across the teams. More details can be found in [14, 15].

We followed two teams. Team Alpha (12 members) is responsible for the app used by travelers. Team Beta (9 members) gathers data from travel companies and structure them into products that other teams use to build their features. We chose these teams because we wanted to explore if coordination strategies differed as Alpha hold lots of dependencies to other teams, while Beta is mostly independent (others are to a large degree dependent on them). We kept an exploratory approach as we did not set out to test any specific theory or hypothesis [16] further, we hold an interpretive view in this study, comprehending the world and its truths as subjective realities [17].

Data collection spanned over three months (November 2021 to January 2022), including eight semi-structured virtual interviews (86 transcribed pages) and notes from two virtual stand-ups. In addition, the first author accessed the virtual workspace of Team Alpha, to observe how members utilized virtual rooms. Analysis was conducted in parallel with data collection, with codes rising inductively from data and forming categories and phenomena. Nvivo was used for coding and building categories. In March 2022, we presented the preliminary findings both in text and in-person presentation to the two teams and facilitated discussions to verify and adjust our findings.

## 4   Results

According to the company guidelines, the teams decided how to execute work-from-anywhere as long as they followed national covid-restrictions. In the period of 24$^{th}$ of September to 30$^{th}$ of November 2021, there were no restrictions. "The offices were completely open, but many choose to use the home office as the main base [in our team]," (B1). Team Alpha came to the office 2-3 days per week, except for a few members that never came in. Team Beta were located in two cities, where three members came to the office most days in one city, while those in the other city rarely went to the office. Prior to the Covid-19 pandemic, all developers in both teams went to the office every day.

### 4.1   Choosing tasks

When choosing tasks from backlogs, developers take their location into consideration – whether they are at home or in the office. While co-located, the teams preferred tasks with an interpretive element, demanding frequent clarifications and discussions. "When developer and designer spend time together – that is the most valuable office-time. […] These tasks have waited about a year, which we pick up now that we are hybrid and back in the office" (A2).

Two criteria are critical when choosing tasks for the home office: One criterion is that the task needs minor clarifications. "I pick simpler tasks [from the backlog] more often for the home office. […] These are just-go-and-do-it tasks that we all agree on how to do," (A1). Informants in both teams tell a similar story of deliberately picking tasks with fewer dependencies with low coordination needs. This way, they "gain a feeling of progression" (B2). Examples of such tasks were bugfixes and small design adjustments.

The second criteria for home tasks is that the task requires uninterrupted focus. "We had this task where everyone worked alone on sub-tasks. We wouldn't gain the same degree of flow if we were at the office, even if we isolated ourselves in a meeting room. Some tasks are best suited when we can isolate at home" (A1). Despite setting up barriers to defend against interruptions, like putting up signs on the meeting room door, co-workers spotted them and found ways to squeeze in a quick talk. It is easier to hide away at home and stay uninterrupted". The team also avoids filling up their calendars with meetings during office days to enable collaborative work. This was a common opinion for all informants.

## 4.2 Use of communication tools

Team Alpha uses tools for mimicking their previous co-located work practices. When the teams were sent home when the pandemic started, an experienced gamer proposed using virtual rooms in Discord to sustain quick clarifications and short exchanges of information the same way online gamers do. They identified several rooms. A "Team-room" imitates their shared space at the office where they all sit together. A room called "One-on-One" imitates meeting rooms where developers can retreat for private discussions. "Do-not-disturb" is like a quiet room.

Observing each other's presence in different rooms provides awareness of coworkers' state of mind. "I can see, for example, that Maria and Peter are sitting in another room and having a meeting. [...] you know where they are [mentally]" (A4). Awareness of what others are doing helps developers interpret if it is appropriate to approach them. "Discord matches how we work when we sit near each other in the office. We can get quick clarifications like 'can you have a brief look at this? Looks OK?'" (A1). Knowing when a person can be contacted lowers the threshold for contacting them, and helps progress in their tasks. All informants in Team Alpha told the same story, often using the same words to describe it.

In contrast, tools like Slack and Teams do not create the same awareness because there is a mistrust of status indicators (indicating i.e. *available* when green and *busy* when red). Unclear statuses make it hard to know when co-workers can be approached/contacted. "You don't know if you are interrupting people when you contact them on Slack. […] you have no idea what they are doing. […] I don't update it [my status indicator] much myself. Based on how I use it myself, I may not fully trust it" (B2). "Yellow or orange or red… I don't dare trust them" (B3). As we have seen, Team Alpha mitigated such challenges by using virtual rooms, while Team Beta relied on Slack.

Implementing tools like Discord requires experimentation. "In the beginning, everyone had their microphone unmuted to make it feel like you were in the office, but at home, you also have other sounds that come from the kitchen or children or cats and stuff, so it did not work well, " (A3). Experimentation led Team Alpha to a prac-

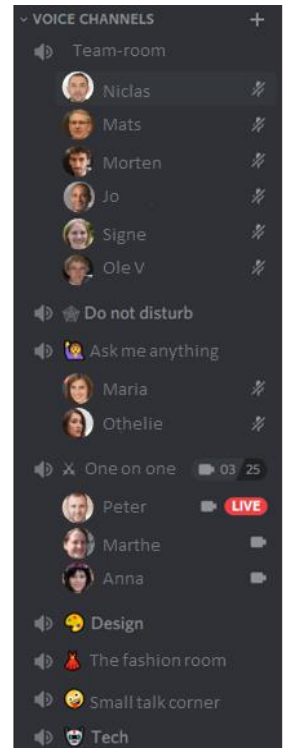

Figure 1 shows the virtual rooms and their participants (pictures are generated by an AI for anonymity). In the 'Team-room', six members are present, all muted but with their speakers on, simulating their shared team space at the office. No one is present in 'Do not disturb'. While two are present in 'Open for questions', they are also muted. Three members have a live discussion in 'One-on-one' with their cameras on. The other rooms, 'Design', 'The Fashion Room', 'Small talk corner', and 'Tech' are empty.

tice where speakers are un-muted, combined with muted microphones when members are not speaking. In that way, they can unmute and ask questions or address someone while everyone hears it. When asked if this is annoying for others in the same virtual room, all informants told us that the practice enabled transparency and opportunities to include oneself. "If you do not like it, you can always turn off your sound, it will be like putting on headphones in the office" (A3). "I thought maybe it would be a little tiring, but it's not. People are very respectful and do not bother each other" (A4).

An important feature is moving members between rooms. "We are all administrators, so that we can move each other between rooms. It's convenient if you want to talk to someone, just enter a room and stick him in there with you and we are off talking. This is the new way of tapping someone on the shoulder when they have their earphones on in the office" (A1).

Although it may be true that virtual rooms maintain unscheduled meetings in virtual settings, things look different on days when the majority of the team is co-located. When presenting preliminary findings to Team Alpha, discussions revealed that they downgraded their use of Discord when coming to the office because they physically observed each other's mental presence. Those few who worked virtually on such days stopped relying on the virtual rooms to communicate teammates' mental presence. However, they all agreed that on non-office days, Discord was still the "lifeline of operations."

### 4.3  Meetings

Unscheduled meetings in the office have transformed into scheduled meetings virtually. Informants highlight this transition as one of the biggest challenges when working virtually. "In the office, it is easy just to say "hey, shall we do this?" and then you have sort of made a clarification in 15 seconds. While digitally, you often end up having to invite for another meeting" (A4). When virtual, people first ask for a talk, then agree if they should meet face to face or virtually, then find a time that suits both calendars. Discord is a way of shortening this process.

While Discord solved the problem of scheduling meetings on team level, the problem still persisted on the inter-team level.: "…each team is on its own Discord server. However, collaboration across teams takes place mainly via Slack or Teams. And there it is again – you have to arrange meetings in advance" (A3).

Even when teams are free to work at the office, inter-team meetings are still challenging. "On those days we were at the office, the other teams weren't" (A3). Informants speculated on various reasons for this: it is more comfortable to go when there are fewer colleagues to share the space with; the best meeting-rooms are available; it is precious time for the teams to meet internally and build cohesion. On the other hand, managers tend to go in on the same days. "Those I need to meet in person [outside my team], I almost always meet them on Tuesdays and Thursdays [their common office days]. Once we have started talking in person, it's easier to take it up again digitally on Slack" (A2).

Interestingly, Team Alpha has concluded that large meetings and retrospectives are exclusively for home-office. The combination of well-functioning virtual white-boards, competition for the best equipped meeting rooms and that teams are seldom present simultaneously makes virtual meetings easier. "There is always someone with a cold or has a sick child, or an [private] appointment to run to. There are always at least two at home" (A2). Virtual meetings led to higher inclusion as everyone always gets to participate. Additionally, retrospectives are automatically documented in virtual whiteboards, whereas they have to convert the whiteboard in physical meetings into digital documents.

## 5    Discussion and conclusion

We have seen how two software development teams over a period of 3 months used various tools and strategies to cope with working from anywhere. Entur offers a full flex solution where teams decide themselves where to work from and how many days at the office. Now, we turn to discuss our research question, *what coordination strategies are used when working from anywhere?* Three distinct strategies that emerge from our data, are summarized in Table 1.

**Table 1.** Strategies for mutual adjustment when working from anywhere

| Strategy | Description/rationale |
| --- | --- |
| Work location decides tasks | Tasks with vague requirements are performed collocated because they often require unscheduled discussions and clarifications, which are more effective in-person. Individual tasks requiring focus are best performed at home. |
| Unscheduled meetings are maintained in virtual rooms | Virtual rooms reveal mental presence to teammates, lowering the threshold for intra-team unscheduled talks. |
| Meeting type decides location | Meetings reporting status are reserved for virtual time to free up office time for unscheduled meetings. Those forced to stay at home, for various reasons, are still included and updated on important information. |

Tasks with vague requirements are chosen for office time because they often require continuous clarifications, joint decision-making, or discussions while working (mutual adjustment or frequent coordination). Our findings are in accordance with Calefato et al. [10] who argue that face-to-face meetings are essential for having in-depth discussions. Co-location seems especially important when tasks require multiple competencies or domains, for example when a developer and designer collaborate on a task. Being co-located makes it easier to adjust to each other's expectations and comprehensions by solving problems together. Further this practice reduced waiting time and blockages which is important for effective coordination [7], and reduced communication problems when solving complex tasks. Teams with communication problems are likely to experience problems coordinating their work [18]. To secure enough time for working co-located, large meetings (typically reporting status) and individual work are down-prioritized, and set aside for the home-office.

Unscheduled meetings are close to the core of mutual adjustment and upheld through virtual rooms. Being present in a room reveals hints about mental presence that help coworkers interpret when it is appropriate to approach them – making it easier to reach out for a quick clarification. For example, when a developer observes a coworker in a meeting room with their manager, he recognizes that this is not the right moment to interrupt. On the other hand, if the developer observes them together at the coffee machine, he can take this opportune moment to interrupt with a quick question. Smite et al. [13] found that a lack of tools showing status of the other teams members was a reason for not being able to mimic the old working practices like pair programming. Further, awareness of what is happening and who is doing what also seemed to initiate unscheduled meetings. Our findings suggest that virtual rooms through Discord facilitates constant informal communication, which improves communication in distributed agile projects [19]. Increased transparency also builds trust, which is vital for distributed teams' success [20].

To conclude, the three strategies affect mutual adjustment by maintaining unscheduled meetings and informal talks. This especially holds true in an intra-team setting, while these strategies seem to struggle in inter-team settings.

Our explorative findings show a need to further understand emerging strategies when WFX. Especially, investigating how these new strategies differ from those already known in the fields of Global Software Engineering and Computer-Supported Cooperative Work (CSCW). Future research should examine the three strategies in new contexts as they will change in the coming years. For example, virtual rooms have only been utilized for a few months in a hybrid setting and will most likely change as teams keep adapting. Also, what long term effects on processes like user involvement, knowledge transfer and onboarding new team members are worth investigating.

**Acknowledgements:** We wish to thank Entur and the informants for willingly sharing their experiences. Also, we thank Knowit AS and the Norwegian Research Council and for funding the research through the projects Transformit (grant number 321477) and A-Team (grant number 267704).